\newcommand{\rem}[1]{}
\documentclass{amsart}
\usepackage{amsfonts,amssymb,amsmath,amsthm,mathrsfs}
\usepackage{url}
\newtheorem{thrm}{Theorem}[section]
\newtheorem{lem}[thrm]{Lemma}
\newtheorem{prop}[thrm]{Proposition}

\newtheorem{remark}[thrm]{Remark}
\theoremstyle{definition}

\author[C.A.Mantica]{Carlo Alberto Mantica}
\address{C. A. Mantica (corresponding author) - I.I.S. Lagrange, Via L. Modignani 65, 20161 Milan,  and I.N.F.N. Sezione di Milano, Via Celoria 16, 20133, Milano, Italy.}
\email{carlo.mantica@mi.infn.it} 
\author[L.G.Molinari]{Luca Guido Molinari}
\address{L. G. Molinari - Physics Department, Universit\`{a} degli Studi di
Milano and I.N.F.N. Sezione di Milano, Via Celoria 16, 20133, Milano, Italy.}
\email{luca.molinari@mi.infn.it}
\author[Y.J.Suh]{Young Jin Suh}
\address{Y. J. Suh - Department of Mathematics \&  RIRCM,  Kyungpook National University,  Daegu 41566,  S. Korea.}
\email{yjsuh@knu.ac.kr}
\author[S.Shenawy]{Sameh Shenawy}
\address{S. Shenawy - Basic Science Department, Modern Academy for
Engineering and Technology, Maadi, Egypt.}
\email{drssshenawy@eng.modern-academy.edu.eg, drshenawy@mail.com}
\begin{document}
\title[Theorems on perfect-fluid and GRW space-times]{PERFECT-FLUID, 
GENERALIZED ROBERTSON-WALKER SPACE-TIMES, And Gray's Decomposition}
\begin{abstract}
We give new necessary and sufficient conditions on the Weyl tensor for generalized Robertson-Walker
(GRW) space-times to be perfect-fluid space-times. For GRW space-times, we determine the form of the Ricci tensor in all the 
O(n)-invariant subspaces provided by Gray's decomposition of the gradient of the Ricci tensor. In all  
but one, the Ricci tensor is Einstein or has the form of perfect fluid. We discuss the corresponding equations of 
state that result from the Einstein equation in dimension 4, where perfect-fluid GRW space-times are Robertson-Walker.
\end{abstract}
\maketitle

\section{Introduction}
Generalized Robertson-Walker (GRW) space-times  
are a natural and wide extension of RW spacetimes, where large scale cosmology is staged. They are 
Lorentzian manifolds of dimension $n$ characterized by the metric
\begin{align}
ds^{2}=-dt^{2}+a^{2}\left( t\right) g_{\mu\nu }^{\ast }\left( \vec{x}%
\right) dx^\mu dx^\nu  \label{E1}
\end{align}%
where $g_{\mu\nu}^{\ast }\left( \vec{x}\right) $ is the metric tensor
of a Riemannian submanifold. A GRW space-time is thus the warped product $%
-I\times _{a}M^*$ where $I$ is an interval of the real line, $\left(
M^{\ast },g^{\ast }\right) $ is a Riemannian manifold and $a>0$ is a smooth warping, or scale function. 
They have been deeply studied in the last years by several authors  
\cite{Alias:1995B}-\cite{Sanchez:1998} (see the review \cite{Mantica:2017C}). 
Few years ago,  Bang-Yen Chen \cite{Chen:2014,Chen:2017}
characterized them by the presence of a time-like concircular vector
in the sense of Fialkow \cite{Fialkow:1939}: 
\begin{thrm}[Chen, 2014]
A Lorentzian manifold  of dimension $n> 3$ is a GRW space-time if and only
if it admits a time-like concircular vector $X_{j}$: 
$X_j X^j<0$ and $\nabla _{k}X_{j}=\rho g_{kj}$, 
where $\rho $ a scalar function.
\end{thrm}
The associated unit time-like vector field $u_{j}= X_j/\sqrt{-X^jX_j}$ turns out to be torse-forming \cite{Yano:1944}:
\begin{align}
\nabla _{j}u_{k}=\varphi \left( g_{jk}+u_{j}u_{k}\right)  \label{E9}
\end{align}
with $\varphi = \rho/\sqrt{-X^jX_j}$. In other words, $u_j$ is a velocity field without shear, vorticity and acceleration. 
The field $\varphi $, in the comoving frame, coincides with Hubble's parameter: $\varphi = \dot a/a$.\\
The alternative characterization was obtained:
\begin{thrm}[Mantica \& Molinari, \cite{Mantica:2017C}]\label{propGRW} 
A Lorentzian manifold of dimension $n> 3$ is a GRW space-time if and only if
it admits a unit time-like torse-forming vector that is also eigenvector of the Ricci
tensor.
\end{thrm}
A further extension are the twisted space-times, where the scale function $a$ may depend on all coordinates.
They were introduced by B.-Y. Chen in 1979 \cite{Chen:1979}, and later
characterized by the existence of a time-like `torqued' vector \cite{Chen_torqued}. 
\begin{thrm}[Mantica \& Molinari, \cite{Mantica:2017B}]\label{thrmtw}
A Lorentzian manifold of dimension $n>3$ is a twisted space-time if and
only if it admits a unit time-like torse-forming vector.
\end{thrm}
A Lorentzian manifold whose Ricci tensor has the form $\mathrm{R}%
_{kl}=A g_{kl}+B v_{k}v_{l}$ with scalar fields $A$, $B$, and a time-like 
`velocity field', $v_{j}v^{j}=-1$, is named
`perfect fluid' space-time \cite{Chen:1979} (a Robertson-Walker space-time is perfect-fluid).
In the geometric literature it is
known as quasi-Einstein manifold 
\cite{Deszcz:1998,Chaki:2000,Shaikh}
(without restriction on $v$). It is an Einstein space-time if ${\mathrm R}_{ij} = ({\mathrm R}/n) g_{ij} $.\\
As $v_j$ is an eigenvector, the Ricci tensor can be parameterized in terms of the scalar curvature 
$\mathrm{R}$ and the eigenvalue $\eta $ as follows:
\begin{align}
\mathrm{R}_{kl}=\frac{\mathrm{R}-n\eta}{n-1}v_k v_l +\frac{\mathrm{R}-\eta }{n-1}g_{kl}  \label{Eperfect}
\end{align}
In section 2, new necessary and sufficient conditions for a GRW space-time to be a perfect fluid, 
and for a perfect fluid to be a GRW space-time, will be given. They are based on the Weyl tensor, 
and extend our result in \cite{Mantica:2016C}. Theorems where the Weyl tensor is replaced by 
other curvature tensors are found in \cite{ManticaOsaka}.

In section 3 we introduce Gray's decomposition \cite{Gray:1978} of the tensor $\nabla_i{\mathrm R}_{jk}$ into
$O(n)$ invariant subspaces, and discuss the special forms of the Ricci tensor of GRW space-times
in each subspace. In all subspaces but one, the Ricci tensor is Einstein or perfect-fluid, with different
restrictions on the scalar curvature and the eigenvalue. They reflect in the equations of state for the
cosmological fluid's pressure and energy density, determined by the Einstein equations, discussed in section 4
for dimension $n=4$. In $n=4$ the perfect fluid GRW space-times coincide with RW space-times.

In the paper the Lorentzian manifolds (space-times) have dimension $n>3$, and are
smooth. When used, a dot means the directional derivative $u^k\nabla_k $.

\section{Perfect-fluid and GRW space-times}
We give new sufficient conditions for a GRW space-time to be perfect-fluid, and for the opposite
occurrence. 
According to Prop.\ref{propGRW}, a GRW space-time is endowed with the special vector $u_j$ \eqref{E9}
that is eigenvector of the Ricci tensor. In Ref.\cite{Mantica:2016A} the 
following general structure of the Ricci tensor was obtained:
\begin{align}
\mathrm{R}_{kl}=\frac{\mathrm{R}-n\xi}{n-1}u_k u_l +\frac{\mathrm{R}-\xi }{n-1}g_{kl} - 
\left( n-2\right) \mathrm{C}_{jklm}u^ju^m  \label{E6}
\end{align}%
where $\mathrm{R}=\mathrm{R}^k{}_k$ denotes the scalar curvature, $\mathrm{C}_{jklm}$ is 
the Weyl tensor and $\xi $ is the 
eigenvalue.

\begin{remark}\label{uv}
Suppose that the GRW space-time is also perfect-fluid, i.e. there is a vector $v_j$ such that
the Ricci tensor has the form \eqref{Eperfect}. Then the condition ${\mathrm R}_{ij}u^j=\xi u_i$ gives
$$\left (\xi  - \frac{\mathrm{R}-\eta }{n-1}\right ) u_k= \frac{\mathrm{R}-n\eta}{n-1}(u^lv_l) v_k $$
Since both $u_k$ and $v_k$ are time-like, it cannot be $u^kv_k=0$. Then, unless the space-time is Einstein, 
it must be  $v_k =\pm u_k$ and $\xi =\eta$.
\end{remark}

We now recall the properties of GRW space-times that are necessary for the discussion. They are mainly
taken from Ref. \cite{Mantica:2016A}.\\
$\bullet$ Unit torse-forming vectors have the property named Weyl compatibility \cite{Mantica:2014A,Mantica_DERDZ}:
\begin{align}
(u_i\mathrm{C}_{jklm}+u_j\mathrm{C}_{kilm}+u_k\mathrm{C}_{ijlm})u^m =0.  \label{E7}
\end{align}
$\bullet$ The eigenvalue is $\xi =(n-1)(\varphi^2 + u^k\nabla_k\varphi )$, 
and $ \nabla_k \xi = -u_k (u^j\nabla_j \xi)$.\\
$\bullet$ The contracted Weyl tensor ${\mathrm C}_{kl} = u^ju^m {\mathrm C}_{jklm} $
has the properties \cite[eqs.14,15]{Mantica:2016A}
\begin{gather}
\nabla^k \mathrm{C}_{kl}=
- \frac{n-3}{2(n-1)(n-2)}(\nabla_l \mathrm{R} + u_l u^k \nabla_k \mathrm{R} ) \label{uC}\\
u^k\nabla_k {\mathrm C}_{jl} = -2\varphi {\mathrm C}_{jl}  \label{recC} 
\end{gather}

The following proposition contains the new statement \eqref{p4}:
\begin{prop}\label{T_equiv}
In a GRW space-time, the following statements are equivalent:
\begin{align}
&  \nabla^m {\mathrm C}_{jklm} =0 \label{p1}\\
&  u^m {\mathrm C}_{jklm} =0  \label{p2}\\
&  {\mathrm C}_{kl} =0  \label{p3} \\
&  u^j \nabla^m {\mathrm C}_{jklm} =0 \label{p4}
\end{align}
\begin{proof}
The equivalence of \eqref{p1} with \eqref{p2} is theorem 3.4 in \cite{Mantica:2016A}.
The equivalence of \eqref{p2} with \eqref{p3} follows from the identity
\begin{align}
 u^m {\mathrm C}_{jklm} = u_k {\mathrm C}_{jl} - u_j {\mathrm C}_{kl} \label{p23}
 \end{align}
which is obtained by contracting \eqref{E7} with $u^j$. Now, \eqref{p3} is equivalent to \eqref{p1} that implies \eqref{p4}.
Let us show that \eqref{p4} implies \eqref{p3}.\\
The covariant divergence of \eqref{p23} 
$(\nabla^j u^m) {\mathrm C}_{jklm} + u^m \nabla^j {\mathrm C}_{jklm} = (\nabla^j u_k)\mathrm{C}_{jl} +
u_k \nabla^j {\mathrm C}_{jl} -(\nabla^j u_j)\mathrm{C}_{kl} -u^j\nabla_j \mathrm{C}_{kl} $ gives:
$ u^m \nabla^j {\mathrm C}_{jklm} = -\varphi (n-1)\mathrm{C}_{kl} +
u_k \nabla^j {\mathrm C}_{jl} -u^j\nabla_j \mathrm{C}_{kl} $. Next use \eqref{recC}:
\begin{align}
u^m \nabla^j {\mathrm C}_{jklm} = -\varphi (n-3)\mathrm{C}_{kl} + u_k \nabla^j {\mathrm C}_{jl} 
\label{unablaC}
\end{align}
If $u^m \nabla^j {\mathrm C}_{jklm}=0$ then $\varphi (n-3)\mathrm{C}_{kl} =
u_k \nabla^j {\mathrm C}_{jl} $. Contraction with $u^k$ gives
$\nabla^j {\mathrm C}_{jl}=0$, but then also ${\mathrm C}_{kl}=0$. 
\end{proof}
\end{prop}
In consideration of the general form \eqref{E6} and of the Remark \ref{uv}, we conclude:
\begin{thrm}\label{T1.6} 
A GRW space-time is perfect fluid if and only if ${\mathrm C}_{kl}=0$,
or any of the equivalent conditions in Prop. \ref{T_equiv}.
\end{thrm}

Now we investigate the problem of a perfect-fluid space-time to be GRW. Namely, given
\begin{align}
\mathrm{R}_{kl}=\frac{\mathrm{R}-n\xi}{n-1}u_k u_l +\frac{\mathrm{R}-\xi }{n-1}g_{kl}  \label{perfect}
\end{align}
we give conditions for the unit time-like vector $u_j$ to be torse-forming. An answer was given with Th 2.1
in \cite{Mantica:2016C}.  Now we extend the result:

\begin{thrm}
A perfect-fluid space-time is GRW if the vector field $u_j$ has the properties: $u^j \nabla^m {\mathrm C}_{jklm}=0$ and
$u^k \nabla_k u_j=0$.
\begin{proof}
The general formula for the divergence of the Weyl tensor is \cite{Postnikov:2001}:
\begin{align}
\nabla_m {\mathrm C}_{jkl}{}^m =\tfrac{n-3}{n-2}\left[\nabla_k {\mathrm R}_{jl} - \nabla_j {\mathrm R}_{kl} -\tfrac{1}{2(n-1)}(g_{jl}\nabla_k \mathrm{R} - g_{kl}\nabla_j \mathrm{R})  \right]  \label{divC}
\end{align}
Contraction with $u^j$ and use of \eqref{perfect} give:
\begin{align*}
0 =&u^j(\nabla_k {\mathrm R}_{jl} - \nabla_j {\mathrm R}_{kl}) -\tfrac{1}{2(n-1)}(u_l\nabla_k \mathrm{R} - g_{kl}u^j \nabla_j \mathrm{R})\\
=& \nabla_k (\xi u_l) - {\mathrm R}_{jl} (\nabla_k u^j) - u^j\nabla_j {\mathrm R}_{kl}  -\tfrac{1}{2(n-1)}(u_l\nabla_k \mathrm{R} - g_{kl}u^j \nabla_j \mathrm{R}) \\
=& \nabla_k (\xi u_l) -\tfrac{\mathrm{R}-n\xi}{n-1}u_ju_l \nabla_k u^l - \tfrac{\mathrm{R}-\xi}{n-1} (\nabla_k u_l) - u^j\nabla_j {\mathrm R}_{kl} \nonumber\\ 
& -\tfrac{1}{2(n-1)}(u_l\nabla_k \mathrm{R} - g_{kl}u^j \nabla_j \mathrm{R}) \\
=&  u_l\nabla_k\xi  - \tfrac{R-n\xi}{n-1} (\nabla_k u_l) - u^j\nabla_j {\mathrm R}_{kl}  -\tfrac{1}{2(n-1)}(u_l\nabla_k \mathrm{R} - g_{kl}u^j \nabla_j \mathrm{R})
\end{align*}
Contraction with $u^l$ and use of $u^j\nabla_k u_j =0$ and $u^j\nabla_j u^l=0$  give:
\begin{align*}
0 &= -\nabla_k \xi  - u^ju^l\nabla_j {\mathrm R}_{kl}  +\tfrac{1}{2(n-1)}(\nabla_k \mathrm{R} +u_ku^j \nabla_j \mathrm{R}) \\
& = -\nabla_k \xi  - u^j\nabla_j (\xi u_k) + {\mathrm R}_{kl}u^j\nabla_j u^l +\tfrac{1}{2(n-1)}(\nabla_k \mathrm{R} +u_ku^j \nabla_j \mathrm{R}) \\
& = -(\nabla_k \xi  + u_k u^j\nabla_j \xi)   +\tfrac{1}{2(n-1)}(\nabla_k \mathrm{R} +u_ku^j \nabla_j \mathrm{R}) 
\end{align*}
The relation $\nabla_k \xi -\tfrac{1}{2(n-1)}\nabla_k \mathrm{R} = -u_k u^j \nabla_j (\xi  -\tfrac{1}{2(n-1)}  \mathrm{R}) $ is inserted back:
\begin{align*}
0 =&  -u_l u_k (u^j \nabla_j \xi ) - \tfrac{\mathrm{R}-n\xi}{n-1} (\nabla_k u_l) - u^j\nabla_j {\mathrm R}_{kl}  +
\tfrac{u^j \nabla_j \mathrm{R}}{2(n-1)}(g_{kl}+u_ku_l) \\
=&  -u_l u_k (u^j \nabla_j \xi ) - \tfrac{\mathrm{R}-n\xi}{n-1} (\nabla_k u_l) - g_{kl} \tfrac{u^j\nabla_j (\mathrm{R}-\xi)}{n-1} -
u_ku_l  \tfrac{u^j\nabla_j(\mathrm{R}-n\xi)}{n-1}\\ 
&+
\tfrac{u^j \nabla_j \mathrm{R}}{2(n-1)}(g_{kl}+u_ku_l) \\
=&  - \tfrac{\mathrm{R}-n\xi}{n-1} (\nabla_k u_l) - (g_{kl} +u_ku_l) u^j\nabla_j \tfrac{\mathrm{R}-2\xi}{2(n-1)}  
\end{align*}
Contraction with $g^{kl}$ gives $
 \tfrac{\mathrm{R}-n\xi}{n-1} (\nabla_k u^k)= -\tfrac{1}{2}u^k\nabla_k (\mathrm{R}-2\xi) $.
Then, if $\mathrm{R}-n\xi \neq 0$, $
\nabla_k u_l = \frac{\nabla_j u^j}{n-1} (g_{kl} + u_k u_l)$, i.e. 
the unit time-like vector is torse-forming. 
\end{proof}
\end{thrm}
The case $\xi =\mathrm{R}/n$ corresponds to an Einstein space-time: ${\mathrm R}_{ij} = 
(\mathrm{R}/n) g_{ij}$.\\
The case $u^k\nabla_k (\mathrm{R}-2\xi )=0$ with $\xi\neq \mathrm{R}/n$, corresponds to $\nabla_k u_l=0$. The space-time
now factors, as the scale factor in \eqref{E1} is trivial ($a=1$).

\section{Gray's decomposition and GRW space-times}
A. Gray \cite{Gray:1978} (see also \cite{Mantica:2012B}\cite[Ch.16]{Besse:2008}) found
that the gradient of the Ricci tensor $\nabla _{j}\mathrm{R}_{kl}$ can be decomposed into $O\left( n\right) $ invariant terms
(see \cite{Hamermesh:1989,Krupka:1995}):
\begin{align}
\nabla_j \mathrm{R}_{kl}=\mathfrak{\mathring{R}}%
_{jkl}+a_j g_{kl}+b_{k}g_{jl}+b_{l}g_{jk}  \label{E11}
\end{align}%
where $\mathfrak{\mathring{R}}_{jkl}=\mathfrak{\mathring{R}}_{jlk}$ is trace-less i.e. $\mathfrak{
\mathring{R}}^j{}_{jk}=\mathfrak{\mathring{R}}_{kj}{}^{j}=0$  and
\begin{align}
a_{j}=\frac{n}{\left( n-1\right) \left( n+2\right) }\nabla _{j}\mathrm{R},%
\text{ \ \ }b_{j}=\frac{n-2}{2\left( n-1\right) \left( n+2\right) }\nabla
_{j}\mathrm{R} \label{coeffs}
\end{align}%
The trace-less tensor can be decomposed as a sum of orthogonal components%
\begin{align}
\mathfrak{\mathring{R}}_{jkl}=\tfrac{1}{3}\left( \mathfrak{\mathring{R}}%
_{jkl}+\mathfrak{\mathring{R}}_{klj}+\mathfrak{\mathring{R}}_{ljk}\right) +%
\tfrac{1}{3}\left( \mathfrak{\mathring{R}}_{jkl}-\mathfrak{\mathring{R}}%
_{kjl}\right) +\tfrac{1}{3}\left( \mathfrak{\mathring{R}}_{jkl}-\mathfrak{%
\mathring{R}}_{ljk}\right)  \label{E12}
\end{align}%
The decomposition \eqref{E11}, \eqref{E12} provides $O\left(
n\right) $ invariant subspaces, characterized by invariant equations that are linear
in $\nabla _{j}\mathrm{R}_{kl}$. In Gray's notation:\\
$\bullet$ The trivial subspace $\nabla _{j}\mathrm{R}_{kl}=0$. \\
$\bullet$ The subspace $\mathcal{I}$  where $\mathfrak{\mathring{R}}_{jkl}=0$, i.e.
\begin{align}
\nabla _{j}\mathrm{R}_{kl}=a_{j}g_{kl}+b_{k}g_{lj}+b_{l}g_{jk}  \label{E13}
\end{align}%
Manifolds satisfying this condition are called Sinyukov manifolds 
\cite{Sinyukova:1979}. \\
$\bullet$ The orthogonal complement $\mathcal{I}^\perp $ where 
$\nabla _{j}\mathrm{R}_{kl}=\mathfrak{\mathring{R}}_{jkl}$ or, equivalently,
$ a_j g_{kl}+b_{k}g_{lj}+b_{l}g_{jk}=0$. 
Then  $\mathcal{I}^\perp $ is only characterized by the equation $\nabla _{j}\mathrm{R}=0$.\\ 
The decomposition \eqref{E12} of $\mathfrak{\mathring{R}}_{jkl}$ specifies orthogonal
subspaces, and $\mathcal{I}^{\perp }=\mathcal{A}\oplus \mathcal{B}\oplus \mathcal{B}'$, where ${\mathcal B}'$ is
a copy of ${\mathcal B}$ with indices exchanged. \\
In $\mathcal{A}$ it is $\mathfrak{\mathring{R}}_{jkl} +\mathfrak{\mathring{R}}_{klj}+\mathfrak{\mathring{R}}_{ljk}=0$ and $\nabla_j \mathrm{R}=0$,  i.e. the Ricci tensor is a Killing tensor \cite{Rani:2003}:
\begin{align}
\nabla _j \mathrm{R}_{kl}+\nabla _k\mathrm{R}_{lj}+\nabla _{l}\mathrm{R}_{jk}=0.  \label{E14}
\end{align}%
In $\mathcal{B}$ and $\mathcal {B'}$ it is $\mathfrak{\mathring{R}}_{jkl} -\mathfrak{\mathring{R}}_{kjl}=0$ and $\nabla_j \mathrm{R}=0$,
i.e. the Ricci tensor is a Codazzi tensor:
\begin{align}
\nabla _{j}\mathrm{R}_{kl}=\nabla _{k}\mathrm{R}_{jl}  \label{E15}
\end{align}%
In all cases the condition $\nabla_j{\mathrm R}=0$ is a 
consequence. Now, we consider two composite subspaces.

The subspace $\mathcal{I}\oplus 
\mathcal{A}$ contains tensors 
that satisfy the cyclic condition%
\begin{align}
\nabla_j  \mathrm{R}_{kl}+\nabla _{k}\mathrm{R}_{jl}+\nabla _{l}\mathrm{R}%
_{kj}=\frac{2\nabla _{j}\mathrm{R}}{n+2}g_{kl}+\frac{2\nabla _{k}\mathrm{R}}{%
n+2}g_{jl}+\frac{2\nabla _{l}\mathrm{R}}{n+2}g_{kj},  \label{E16}
\end{align}%
i.e. the Ricci tensor is a conformal Killing tensor \cite{Rani:2003}. Note that the cyclic sum
of \eqref{E13} gives \eqref{E16} (the Ricci tensor of a Sinyukov manifold
is conformal Killing).

 The subspace $\mathcal{I}\oplus \mathcal{B}$ contains tensors 
that satisfy the Codazzi condition%
\begin{align}
\nabla_j \left[ \mathrm{R}_{kl}-\frac{\mathrm{R}}{2\left( n-1\right) }g_{kl}\right ]=
\nabla_k \left[ \mathrm{R}_{jl}-\frac{\mathrm{R}}{2\left(
n-1\right) }g_{jl} \right]. \label{E17}
\end{align}%

Manifolds satisfying conditions \eqref{E13}-\eqref{E17} are also called ``Einstein-like manifolds'' 
(see \cite{Mantica:2017A} and references therein).\\

It is interesting to find the form of the Ricci tensor 
of GRW space-times in Gray's subspaces.
The gradient of the Ricci tensor 
and the divergence of the Weyl tensor are linked by the identity \eqref{divC}, which becomes:
\begin{align}
\nabla_m {\mathrm C}_{jkl}{}^m = 
 -\frac{n-3}{n-2} (\mathfrak{\mathring{R}}_{jkl} -\mathfrak{\mathring{R}}_{kjl} ) \label{nablaCbis}
\end{align}

\subsection{Ricci tensor in the trivial subspace} If $\nabla_j \mathrm{R}_{kl}=0$ the gradient of
$R_{kl}u^l=\xi u_k$ gives $\mathrm{R}_{kl}=\xi g_{kl}$: the GRW space-time is Einstein.

\subsection{Ricci tensor in the subspace $\mathcal I$}
The Ricci tensor in the subspace $\mathcal{I}$ satisfies the condition $\mathfrak{\mathring{R}}_{jkl}=0$ or 
\eqref{E13}. Eq.\eqref{nablaCbis}
shows that a Sinyukov manifolds is perfect fluid (quasi-Einstein). 

\begin{lem}
If the tensor $\alpha_j g_{kl} + \beta_k g_{lj} + \gamma_l g_{jk} +\delta_j v_kv_l$, with $v^2=v^kv_k \neq 0$, 
is zero, then the vector coefficients are zero, $\alpha_i=\beta_i=\gamma_i=\delta_i=0$.
\begin{proof}
Contraction with $v^k$ gives $\alpha_j v_l + (\beta_k v^k)g_{lj} + \gamma_l v_j + v^2 \delta_j v_l=0$.
Contraction with $\tau^l\tau^j$, with $\tau^k v_k=0$ and $\tau^2\neq 0$ gives $\beta_k v^k=0$. Then
$(\alpha_j +\delta_j v^2)v_l + \gamma_l v_j =0$ i.e. both $\gamma_l$ and
$(\alpha_l+\delta_l v^2)$ are parallel to $v_l$.\\
Similarly, contraction with $v^l$ and then with $\tau^j\tau^k$ gives $\gamma_lv^l=0$ and both $\beta_l$ and
$(\alpha_l +\delta_l v^2)$ parallel to $v_l$. Now, $\beta_l$ cannot be orthogonal and parallel to $v_l$,
and the same for $\gamma_l$. Then $\beta_l=\gamma_l=0$.
Next, consider $\alpha_j g_{kl} +\delta_j v_kv_l =0$. Contraction with $\tau^k$  gives $\alpha_j=0$, 
and then $\delta_j=0$.
\end{proof}
\end{lem}
\begin{thrm}\label{TSyn}
The Ricci tensor of a GRW space-time belongs to $\mathcal{I}$ if and only if the space-time is perfect fluid and $(n+2)\dot\xi = 2 \dot{\mathrm R}$. In the comoving frame it is 
\begin{align}
 \mathrm{R} (t)= \alpha - \beta \frac{n+2}{n-2}a(t)^2, \qquad \xi (t) = 
 \frac{\alpha}{n} - \beta \frac{2}{n-2} a(t)^2 \label{solution_I}
 \end{align}
where $\alpha$ and $\beta $ are constants and $a(t)$ is the warping function.
\begin{proof}
If the Ricci tensor belongs to $\mathcal I$, then \eqref{E13} and \eqref{coeffs} give 
$\nabla_m \mathrm{C}_{jkl}{}^{m}=0$, the Ricci tensor has 
the perfect fluid form \eqref{perfect}, and $\nabla_j {\mathrm R} = -u_j \dot {\mathrm R}$. 
Now, evaluate:
\begin{align*}
\nabla_j {\mathrm R}_{kl} = u_k u_l (2\varphi u_j + \nabla_j) \frac{\mathrm{R}-n\xi}{n-1} + g_{kl} \nabla_j \frac{\mathrm{R}-\xi}{n-1} + \varphi\frac{\mathrm{R}-n\xi}{n-1} (u_l g_{jk}+ u_k g_{jl})
\end{align*}
By subtracting \eqref{E13} one obtains a null tensor which, by the previous lemma, 
implies the constraints:
\begin{align*}
& \nabla_j (\mathrm{R}-n\xi) = -2\varphi u_j (\mathrm{R}-n\xi), \\  
& (n+2)\nabla_j(\mathrm{R}-\xi)= n\nabla_j \mathrm{R}, \\  
& 2(n+2)\varphi (\mathrm{R}-n\xi) u_j = (n-2)\nabla_j \mathrm{R}
\end{align*}
The system is supplied with the equation resulting from the covariant derivative of  ${\mathrm R}_{kl} u^l =\xi u_k$:
\begin{align}
\dot {\mathrm R} -2 \dot\xi = -2\varphi (\mathrm{R}-n\xi). \label{Rxi}
\end{align}
The system is degenerate and has solution
\begin{align}
\dot {\mathrm R} = \frac{n+2}{2}\dot \xi , \qquad  \dot {\mathrm R} -n\dot\xi = 2\varphi (R-n\xi ). \label{twoeqs} 
\end{align} 
The equations \eqref{twoeqs} can be integrated. In the comoving frame it is $\varphi = \dot a/a$, then 
the second equation yields $\mathrm{R}-n\xi = \beta a(t)^2$, where
$\beta $ is a constant an $a(t)$ is the warping function. 
The first equation is now used: $\dot {\mathrm R} = -2\beta\frac{n+2}{n-2} a \dot a$, and the results are obtained.

On the other hand, suppose that the GRW space-time is perfect fluid, and
that $(n+2)\dot\xi = 2\dot{\mathrm R}$ holds.
The gradient of the Ricci tensor \eqref{perfect} is
\begin{align*}
\nabla_j{\mathrm R}_{kl} =& \tfrac{1}{n-1}[\nabla_j ({\mathrm R} -n\xi) + 2
\varphi u_j ({\mathrm R}-n\xi)] u_k u_l \\
&+\tfrac{1}{n-1}[\nabla_j ({\mathrm R}-\xi) g_{kl} + \varphi ({\mathrm R}-n\xi)(
u_k g_{jl} + u_l g_{jk})]
\end{align*} 
The first term is zero because for a GRW perfect fluid: $\nabla_j ({\mathrm R} -n\xi) =
-u_j(\dot {\mathrm R} -n\dot \xi) = -2\varphi u_j ({\mathrm R} -n\xi)$ by \eqref{Rxi} and 
$(n+2)\dot\xi = 2\dot{\mathrm R}$. Then:
\begin{align}
\nabla_j{\mathrm R}_{kl} =& \tfrac{1}{n-1}[-u_j (\dot{\mathrm R}-\dot \xi) g_{kl} + \tfrac{1}{2}
(\dot {\mathrm R}-n\dot \xi)(u_k g_{jl} + u_l g_{jk})]  \nonumber\\
=&- \tfrac{1}{(n-1)(n+2)} \dot{\mathrm R} [nu_jg_{kl} +\tfrac{n-2}{2}(u_k g_{jl} + u_l g_{jk})] \nonumber\\
=& \tfrac{n}{(n-1)(n+2)} g_{kl}\nabla_j{\mathrm R} +\tfrac{n-2}{2(n-1)(n+2)}(g_{jl}\nabla_k{\mathrm R} 
+ g_{jk}{\mathrm R}) \label{RSY}
\end{align} 
which is the Sinyukov condition \eqref{E13}.
\end{proof}
\end{thrm}

\subsection{Ricci tensor in the subspace $\mathcal A$}
The subspace $\mathcal{A}$ is characterized by the condition $\nabla_k \mathrm{R}_{ij}+\nabla _{i}
\mathrm{R}_{kj}+\nabla _{j}\mathrm{R}_{ki}=0$, giving
$\nabla_k\mathrm{R}=0$. We now show that (in a GRW space-time) it is $\nabla_k R_{jl}=0$. The subspace is then empty, as the case is accounted for by the trivial subspace. 

\begin{thrm}
In the subspace $\mathcal A$ the Ricci tensor is Einstein.
\begin{proof}
Contraction with $u^j$ of \eqref{E14} gives:
\begin{align*}
0=&u^j\nabla_j {\mathrm R}_{kl}+\nabla_k ({\mathrm R}_{lj}u^j)-{\mathrm R}_{lj}\nabla_k u^j + \nabla_l({\mathrm R}_{jk}u^j) - {\mathrm R}_{jk} \nabla_l u^j\\
=&\dot {\mathrm R}_{kl} +\nabla_k (\xi u_l)-\varphi {\mathrm R}_{lj}(u_k u^j+\delta_k^j) + \nabla_l(\xi u_k) -\varphi {\mathrm R}_{jk} (u_l u^j+\delta_l^j)\\
=&\dot {\mathrm R}_{kl} + (u_l\nabla_k\xi +u_k\nabla_l\xi)  -2\varphi ({\mathrm R}_{kl} - \xi g_{kl}) 
\end{align*}
Next, use $\nabla_k\xi =-u_k\dot \xi$ to obtain:
\begin{align}
0=u^p\nabla_p {\mathrm R}_{kl} -2 u_l u_k\dot \xi  -2\varphi ({\mathrm R}_{kl} - \xi g_{kl})  \label{dotRicci}
\end{align}
Contraction with  $u^k$ gives: $0=3u_l\dot\xi $ i.e. $\dot\xi =0$. On the other hand, contraction with $g^{kl}$ gives 
$\dot {\mathrm R}+2\dot\xi = 2\varphi ({\mathrm R}-n \xi)$. Since $\dot {\mathrm R}=0$ and $\dot\xi=0$ it is 
${\mathrm R}=n\xi$. Then:
\begin{align*}
\mathrm{R}_{kl} = \frac{\mathrm R}{n} g_{kl} - (n-2) {\mathrm C}_{kl}   
\end{align*}
If this is inserted in \eqref{dotRicci}, we obtain $u^p\nabla_p {\mathrm C}_{kl} =2\varphi {\mathrm C}_{kl} $. This is 
in contrast with \eqref{recC}, unless ${\mathrm C}_{kl} =0$. 
\end{proof}
\end{thrm}

\subsection{Ricci tensor in the subspace $\mathcal B$}
In this subspace the Ricci tensor is Codazzi, \eqref{E15}. A contraction
with the metric tensor gives $\nabla _{k}\mathrm{R}=0$, and \eqref{divC}
gives $\nabla_m \mathrm{C}_{jkl}{}^m=0$. Therefore, the GRW space-time is
perfect fluid.\\ 
The equation $\dot\xi = \varphi (R-n\xi)$ can be integrated: in the
comoving frame, where $\varphi = \dot a/a$, the eigenvalue depends on time through the
warping function as $\xi (t) = \alpha a(t)^{-n} + {\mathrm R}/n$, with constant $\alpha$.

\subsection{Ricci tensor in the subspace ${\mathcal I}^\perp$} In this case
$\nabla_k {\mathrm R}=0$. The GRW space-time is not in general perfect-fluid, with the Ricci
tensor having the form \eqref{E6}. However, the equation $\dot\xi = \varphi (R-n\xi)$ can be integrated
and $\xi (t) = \alpha a(t)^{-n} + {\mathrm R}/n$, with constant $\alpha$.


\subsection{Ricci tensor in the subspace $\mathcal{I}\oplus \mathcal{B}$}
The Ricci tensor satisfies the Codazzi condition \eqref{E17}, which is
necessary and sufficient for the divergence of the Weyl tensor \eqref{divC} 
to vanish. Therefore, the Ricci tensor has the perfect fluid form \eqref{perfect}.

\subsection{Ricci tensor in the subspace $\mathcal{I}\oplus \mathcal{A}$}
In this subspace the Ricci tensor is conformal Killing, \eqref{E16} (see \cite{Sharma_Gosh}). 
We now show that the subspaces $\mathcal I\oplus \mathcal A$ and $\mathcal I$ coincide. 


%

\begin{thrm}\label{T5.1} 
The Ricci tensor in a GRW space-time is conformal Killing if and only if it has the perfect fluid form
and $(n+2)\dot\xi = 2 \dot{\mathrm R}$, i.e. it belongs to $\mathcal I$.
\begin{proof}
Suppose that the Ricci tensor is conformal Killing. On multiplying \eqref{E16} by $u^ju^k$ we get%
\begin{align*}
2u^j\nabla_j (u^k \mathrm{R}_{kl})+u^ju^k\nabla_l\mathrm{R}_{kj}=\tfrac{4}{n+2}
\dot{\mathrm R} u_l -\tfrac{2}{n+2}\nabla _{l}\mathrm{R}.
\end{align*}
It is $u^ju^k\nabla_l\mathrm{R}_{kj}=u^j\nabla_l(\xi u_j)-u^j\mathrm{R}_{kj}\nabla_l u^k =
-\nabla_l\xi -\xi u_k \nabla_l u^k = -\nabla_l\xi =u_l\dot\xi$. Then:
\begin{align*}
3u_l \dot\xi  =\tfrac{4}{n+2}\dot{\mathrm R}u_l - \tfrac{2}{n+2} \nabla _l\mathrm{R}.
\end{align*}
Contraction with $u^l$ gives $(n+2) \dot \xi =2 \dot{\mathrm R}$. This, when inserted back, gives
$\nabla_k R = -\dot {\mathrm R} u_k $ and, because of \eqref{uC}: $\nabla_k {\mathrm C}^k{}_l=0$. 
The general property \eqref{Rxi}, $\dot {\mathrm R} -2 \dot\xi = -2\varphi (R-n\xi)$ gives:
\begin{align}
\dot{\mathrm R} -n\dot \xi= 2\varphi (\mathrm{R}-n\xi ).  \label{5E3}
\end{align}%
Transvect \eqref{E16} by $u^j$ and simplify with $\nabla_k \mathrm{R} =-\dot {\mathrm R} u_k$ and the identity
 $u^j \nabla_k {\mathrm R}_{jl} = \nabla_k (\xi u_l) - {\mathrm R}_{jl}\nabla_k u^j =
-\dot\xi u_ku_l +\varphi \xi g_{kl} -\varphi {\mathrm R}_{kl}$
\begin{align*}
u^j\nabla_j  \mathrm{R}_{kl}=&-u^j\nabla _{k}\mathrm{R}_{jl}-u^j\nabla _{l}\mathrm{R}_{kj}
+\tfrac{2}{n+2}\dot{\mathrm R} (g_{kl}-2u_ku_l) \nonumber\\
=&\, 2\dot\xi u_ku_l -2\varphi \xi g_{kl} +2\varphi {\mathrm R}_{kl}+\tfrac{2}{n+2}\dot{\mathrm R} (g_{kl}-2u_ku_l).
\end{align*}
Now use $(n+2)\dot\xi =2\dot {\mathrm R}$ and obtain: $
u^j\nabla_j \mathrm{R}_{kl}= 2\varphi {\mathrm R}_{kl}-2\varphi \xi g_{kl} +\dot \xi g_{kl}$.
The left-hand side of the equation is now evaluated with \eqref{E6}, with the aid of \eqref{recC}:
\begin{align*}
u^j\nabla_j \mathrm{R}_{kl}=& \frac{\dot {\mathrm R} - n\dot \xi}{n-1}u_ku_l + \frac{\dot {\mathrm R} -\dot \xi}{n-1}g_{kl} 
-(n-2) u^j\nabla_j {\mathrm C}_{kl}\\
=& 2\varphi \frac{\mathrm{R} - n\xi}{n-1}u_ku_l + 2\varphi \frac{\mathrm{R} -n\xi}{n-1}g_{kl} +\dot\xi g_{kl}
+2\varphi (n-2) {\mathrm C}_{kl}\\
=& 2\varphi {\mathrm R}_{kl} -2\varphi \xi g_{kl} + \dot\xi g_{kl}+4\varphi (n-2) {\mathrm C}_{kl}.
\end{align*}
This and the previous equation imply ${\mathrm C}_{kl}=0$. Then the GRW space-time is 
perfect-fluid with $(n+2)\dot\xi = 2 \dot{\mathrm R}$.\\
The proof of the opposite statement runs as for theorem \ref{TSyn}, and obtains \eqref{RSY}. A cyclic summation gives that the Ricci tensor is conformal Killing.
\end{proof}
\end{thrm}


\section{Perfect-fluid equations of state}
We examine the equations of state that arise from the perfect-fluid solutions in Gray's 
subspaces. The perfect-fluid form of the Ricci tensor corresponds, via the Einstein equations
${\mathrm R}_{ij}-\tfrac{1}{2}Rg_{ij} = \kappa T_{ij}$
to a perfect fluid energy-momentum tensor $T_{ij} = (p+\mu) u_iu_j + p g_{ij}$.
By assuming the expression \eqref{perfect} for the Ricci tensor, the Einstein equation gives the pressure and the energy-density in terms of $\mathrm R$ and $\xi$: 
\begin{align}
 \kappa p = \tfrac{1}{n-1}({\mathrm R-\xi}) - \tfrac{1}{2}{\mathrm R}, \qquad \kappa\mu =
\tfrac{1}{2}{\mathrm R}-\xi \label{pmu}
\end{align}

We recall Proposition 3.1 in \cite{Mantica:2016C}: a perfect fluid 
space-time in dimension $n\geq 4$ with
differentiable state equation $p=p\left( \mu \right)$, $p+\mu \neq 0$ and
with null divergence of the Weyl tensor $\nabla _{m}\mathrm{C}_{ijk}{}^{m}=0$
is a GRW space-time. Null divergence implies that the Ricci tensor belongs to
the subspace $\mathcal{I\oplus B}$. On the other hand, consider a
perfect fluid GRW space-time with state equation $p=\frac{\mu }{n+1}+\mathrm{constant}$: it
follows that $\nabla _{k}R=0$. This and  $\nabla _{m}\mathrm{C}_{ijk}{}^{m}=0$ give that the Ricci tensor is Codazzi, 
i.e. it belongs to $\mathcal{B}$. Then a perfect fluid GRW space-time with a
state equation different from $p=\frac{\mu }{n+1}+\mathrm{constant}$ 
belongs to $\mathcal{I\oplus B}$ and not to $\mathcal{B}$.\\

Hereafter, we restrict to dimension $n=4$, where a GRW perfect-fluid space-time is exactly a
Robertson-Walker (RW) space-time (this follows from $u^m{\mathrm C}_{jklm}=0$ which, in $n=4$, 
is equivalent to $u_i{\mathrm C}_{jklm} + u_j{\mathrm C}_{kilm} +u_k{\mathrm C}_{ijlm} =0$ as
shown in Lovelock and Rund, \cite{Lovelock} page 128. Contraction with $u^i$ gives 
${\mathrm C}_{jklm}=0$).\\
The cases are:\\

$\bullet$ In the trivial subspace the space-time is Einstein $({\mathrm R}=4\xi$, 
$p=-\mu $).\\

$\bullet$ In $\mathcal B$ the Ricci tensor is Codazzi and the RW space-time
is a ``Yang's Pure Space'' \cite{Guilfole}. Since ${\mathrm R}$ is constant, eq.\eqref{pmu}
gives the equation of state $p=\frac{1}{3}\mu -\frac{1}{3\kappa} {\mathrm R}$. \\
The RW spaces with constant ${\mathrm R}$ are described, for example, \cite{Melia}. In the expanding ones, 
the time evolution of the eigenvalue 
$\xi (t) = \alpha a(t)^{-4} + \tfrac{1}{4}{\mathrm R}$
drives the space-time to an Einstein space-time with $\kappa\mu_\infty ={\mathrm R}/4$ and negative pressure $\kappa p_\infty = -{\mathrm R}/4$. 
Then, asimptotically in the future $p=-\mu$. The case with spatial curvature $R^*=0$ and its cosmological implications are studied 
in \cite{MolManCosmol}.\\

$\bullet$ In $\mathcal I$ 
the solution \eqref{solution_I} gives the dependence in the cosmological time of pressure and density, via the warping function:
\begin{align}
\kappa p(t) = -\tfrac{1}{4}\alpha   + \tfrac{5}{6} \beta  a(t)^2\quad  \kappa\mu (t)= \tfrac{1}{4}\alpha -\tfrac{1}{2}\beta
a(t)^2
\end{align}
Elimination of $\beta a(t)^2$ gives a phantom-type equation of state $p = -\frac{5}{3} \mu+ \rm{const}$,
studied by Caldwell \cite{Caldwell}.\\

$\bullet$ In $\mathcal I \oplus \mathcal B$ the RW space-time is unrestricted.\\

$\bullet$ In ${\mathcal I}^\perp $, the Ricci tensor contains the Weyl term, then it is not perfect-fluid, and the GRW
space-time is not RW. However,
the condition $\nabla_k {\mathrm R}=0$ gives the time evolution
of the eigenvalue of the Ricci tensor $\xi (t) = \alpha a(t)^{-4} + \frac{1}{4}{\mathrm R}$.


\begin{table}
\begin{center}
\begin{tabular}{ |c | l l |c| l | }   
\hline
Subsp. & Condition on $\nabla _{j}{\mathrm R}_{kl}$ & {} & {} & $p(\mu)$, $n=4$\\ 
\hline
Trivial & $\nabla _{j}{\mathrm R}_{kl}=0$ & Ricci symmetric & E &$p=-\mu $\\[4pt]
$\mathcal{I}$ & $\nabla _{j}\mathrm{R}_{kl}=a_{j}g_{kl}+b_{k}g_{lj}+b_{l}g_{jk}$ 
& Sinyukov & pf & $p=-\tfrac{5}{3}\mu + {\rm c}$ \\ [4pt]
$\mathcal{A}$ & $\nabla_{(j}\mathrm{R}_{kl)}=0$ & Killing & $\emptyset $& -- \\ [4pt]
$\mathcal{B}$ & $\nabla _{[j}{\mathrm R}_{k]l}=0 $ & Codazzi & pf & $p=\tfrac{1}{3}\mu+\mathrm c $ \\[4pt] 
$\mathcal{I\oplus A}$ &  $\nabla _{(j}\mathrm{R}_{kl)} = \frac{2}{n+2} \nabla _{(j}\mathrm{R} g_{kl)} $  & Conformal Killing 
& pf & $p=-\tfrac{5}{3}\mu + {\rm c}$ \\ [4pt]
$\mathcal{I\oplus B}$ & $\nabla_{[j} \mathrm{R}_{k]l} = \frac{1}{2 (n+2) }\nabla_{[j}\mathrm{R} g_{k]l} $ & 
$\mathfrak{\mathring{R}}_{jkl}=\mathfrak{\mathring{R}}_{kjl}$ & pf & unrestricted \\ [4pt]
$\mathcal{I}^{\perp }$ & $\nabla _{j}\mathrm{R}=0$ & Const. scalar curv. & -- & --  \\
\hline
\end{tabular}
\caption{GRW space-times in Gray's decomposition (E=Einstein, pf=perfect fluid, $T_{[jk]} = T_{jk}-T_{kj}$, $T_{(jkl)} =T_{jkl}+T_{klj}+ T_{ljk}$, c is a constant).} 
\label{Table:1}
\end{center}
\end{table}

\section*{Acknowledgments}
The third author was supported by grant Proj. NRF-2018-R1D1A1B-05040381 
from the National Research Foundation of Korea. 

We thank the referee for his valuable suggestions, that helped us in improving and 
clarifying the final form of the paper.


\begin{thebibliography}{99}
%
\bibitem{Alias:1995B} 
L. Al{\'\i}as, A. Romero, and M. S\'anchez,
\emph{Uniqueness of complete spacelike hypersurfaces of constant mean curvature in
generalized Robertson-Walker space-times}, Gen. Relativ. Gravit. {\bf 27}(1),
(1995), 71-84.
%
\bibitem{Brozos:2005} M. Brozos-V\'{a}zquez, E. Garcia-Rio, and R.
Vazquez-Lorenzo, \emph{Some remarks on locally conformally flat static
space-times}, J. Math. Phys. {\bf 46} (2005), 022501.
%
\bibitem{Mantica:2016C} C. A. Mantica, L. G. Molinari and U. C. De, \emph{A
condition for a perfect fluid space-time to be a generalized
Robertson-Walker space-time}, J. Math. Phys. {\bf 57} (2) (2016), 022508. Erratum
57, 022508 (2016).
%
\bibitem{Mantica:2016B} 
C. A. Mantica , Y. J. Suh, and U. C. De, 
\emph{A note on generalized Robertson-Walker space-times}, 
Int. J. Geom. Meth. Mod. Phys. {\bf 13} (2016), 1650079, 9 pp.
%
\bibitem{Mantica:2016A} 
C. A. Mantica and L. G. Molinari, 
\emph{On the Weyl and the Ricci tensors of Generalized Robertson-Walker space-times}, 
J. Math. Phys. {\bf 57} (10) (2016), 102502.
%
\bibitem{Romero:2013B}
 A. Romero, R. N. Rubio and J. J. Salamanca, 
\emph{Uniqueness of complete maximal hypersurfaces in spatially parabolic
generalized Robertson-Walker space-times}, Class. Quantum Grav. {\bf 30} (11)
(2013), 115007.
%
\bibitem{Sanchez:1999} 
M. S\'anchez, \emph{On the geometry of generalized Robertson-Walker spacetimes: Curvature and Killing fields}. 
Gen. Relativ. Gravit. {\bf 31} (1999), 1--15.
%
\bibitem{Sanchez:1998} 
M. S\'anchez, \emph{On the geometry of generalized Robertson-Walker spacetimes: geodesics}, 
Gen. Relativ. Gravit. {\bf 30} (1998), 915--932.
%
\bibitem{Mantica:2017C} 
C. A. Mantica and L. G. Molinari, 
\emph{Generalized Robertson-Walker space-times, a survey}, 
Int. J. Geom. Meth. Mod. Phys. {\bf 14} (3) (2017) 1730001 (27pp).
%
\bibitem{Chen:2014} 
B.-Y. Chen, \emph{A simple characterization of generalized Robertson-Walker space-times}, 
Gen. Relativ. Gravit. {\bf 46} (2014), 1833 (5pp).
%
\bibitem{Chen:2017} 
B.-Y. Chen, \emph{Differential geometry of warped product manifolds and submanifolds}, World Scientific (2017).
%
\bibitem{Fialkow:1939} 
A. Fialkow, \emph{Conformal geodesics}, Trans. Amer.
Math. Soc. {\bf 45} (3) (1939), 443--473.
%
\bibitem{Yano:1944} 
K. Yano, \emph{On the torseforming direction in Riemannian Spaces}, 
Proc. Imp. Acad. Tokyo {\bf 20} (1944), 340--345.
%
\bibitem{Chen:1979} 
B.-Y. Chen, \emph{Totally umbilical submanifolds},
Soochow J. Math. {\bf 5} (1979), 9--37.
%
\bibitem{Chen_torqued}
B.-Y. Chen, {\em Rectifying submanifolds of Riemannian manifolds and torqued vector fields}, Kragujev.
J. Math. {\bf 41} (1) (2017), 93Ð103.
%
\bibitem{Mantica:2017B} 
C. A. Mantica and L. G. Molinari, 
\emph{Twisted Lorentzian manifolds: a characterization with torse-forming time-like unit
vectors}, Gen. Relativ. Gravit. 49:51 (2017).
%
\bibitem{Deszcz:1998} 
R. Deszcz, M. G{\l}ogowska, M. Hotlo\'s and Z. Sent\"urk, 
\emph{On certain quasi-Einstein semisymmetric hypersurfaces}, 
Annu. Univ. Sci. Budapest Eotvos Sect. Math. {\bf 41} (1998), 151--164.
%
\bibitem{Chaki:2000} M. C. Chaki and R.K. Maity 
\emph{On quasi Einstein manifolds}, 
Publ. Math. Debrecen {\bf 57} (2000), 257--306.
%
\bibitem{Shaikh}
A. A. Shaikh, D. W. Yoon, S. K. Hiu, {\em On quasi-Einstein space-times}, Tsukuba J. Math. {\bf 33} (2)
(2009) 305--326.
%
\bibitem{ManticaOsaka}
C. A. Mantica, U. C. De, Y. J. Suh and L. G. Molinari,
{\em Perfect fluid spacetimes with harmonic generalized curvature tensor},
Osaka J. Math. {\bf 56} (2019) 173--182.
%
\bibitem{Gray:1978} 
A. Gray,  \emph{Einstein-like manifolds which are not Einstein}, Geom. Dedicata {\bf 7} (1978), 259--280.
%
\bibitem{Mantica:2014A} 
C. A. Mantica and L. G. Molinari, 
\emph{Weyl compatible tensors}, Int. J. Geom. Meth. Mod. Phys. {\bf 11} (8) (2014), 1450070.
%
\bibitem{Mantica_DERDZ} 
C. A. Mantica and L. G. Molinari, 
\emph{Extended Derdzi\'nski-Shen theorem for curvature tensors}, Colloq. Math. {\bf 128} (1) (2012), 1--6.
%
\bibitem{Postnikov:2001} M. M. Postnikov, \emph{Geometry VI, Riemannian
geometry}, Encyclopaedia of Mathematical Sciences, Vol. 91, 2001,
Springer-Verlag, Berlin (translation of 1998 Russian edition by S.A. Vakhrameev).
%
\bibitem{Mantica:2012B} 
C. A. Mantica and L. G. Molinari, \emph{Riemann compatible tensors}, Colloq. Math. {\bf 128} (2012), 197--210.
%
\bibitem{Besse:2008} 
A. L. Besse, \emph{Einstein Manifolds}, Classics in Mathematics, Springer-Verlag, Berlin, (2008).
%
\bibitem{Hamermesh:1989} 
M. Hamermesh, \emph{Group Theory and its Application to Physical Problems}, Dover (1989).
%
\bibitem{Krupka:1995} 
D. Krupka, \emph{The trace decomposition problem}, Beitr\"age Algebra Geom. {\bf 36} (1995), 303--315.
%
\bibitem{Sinyukova:1979} N. S. Sinyukov, 
\emph{Geodesic Mappings of Riemannian Spaces}, Nauka, Moscow (1979) (in Russian)
%
\bibitem{Rani:2003} R. Rani, B. Edgar, and A. Barnes, \emph{Killing tensors
and conformal Killing tensors from conformal Killing vectors}, 
Class. Quantum Grav. {\bf 20} (11) (2003), 1929--1942.
%
\bibitem{Mantica:2017A} 
C. A. Mantica and S. Shenawy, \emph{Einstein-like
warped product manifolds}, Int. J. Geom. Meth. Mod. Phys. {\bf 14} (11) (2017), 1750166 (11pp.)
%
\bibitem{Sharma_Gosh}
R. Sharma and A. Ghosh, {\emph Perfect-fluid space-times whose energy-momentum tensor is conformal Killing},
J. Math. Phys. {\bf 51} (2010) 022504.
%
\bibitem{Lovelock}
D. Lovelock and H. Rund, \emph{Tensors, Differential Forms and Variational Principles}, 
Reprinted Edition (Dover, 1988).
%
\bibitem{Guilfole}
B. S. Guilfoyle and B. C. Nolan, \emph{Yang's gravitational theory}, Gen. Relativ. Gravit.
{\bf 30} (3) (1998) 473--495.
%
\bibitem{Melia}
F. Melia, {\em Cosmological redshift in Friedmann-Robertson-Walker metrics with constant space-time curvature},
Mon. Not. R. Astron. Soc. {\bf 422} (2012) 1418--1424.
%
\bibitem{MolManCosmol}
L. G. Molinari and C. A. Mantica, {\em w=1/3 to w=-1 evolution in Robertson-Walker space-times with constant scalar curvature}, 
Int. J. Geom. Meth. Mod. Phys. {\bf 16} (2019) 1950061 (9pp.)
%
\bibitem{Caldwell}
R. R. Caldwell, {\em A Phantom Menace? Cosmological consequences of a dark energy component with 
super-negative equation of state}, Phys. Lett. B {\bf 545} (2002) 23--29.
%
\end{thebibliography}
\end{document}